\pdfoutput=1
\documentclass[11pt]{article}

\usepackage[]{acl}

\usepackage{times}
\usepackage{latexsym}
\usepackage[T1]{fontenc}
\usepackage[utf8]{inputenc}
\usepackage{microtype}
\usepackage{graphicx}
\usepackage{booktabs}
\usepackage{enumitem}

\newcommand{\datasetname}{CLAP\textsc{nq}}{}
\newcommand{\system}{\textsc{InspectorRAGet}}{}

\newcommand{\screenshot}[1]{\ref{fig:combo.interface}#1}

\title{\system: An Introspection Platform for RAG Evaluation}
\author{
    Kshitij Fadnis \\ {\footnotesize kpfadnis@us.ibm.com}   \And
    Siva Sankalp Patel\\ {\footnotesize siva.sankalp.patel@ibm.com} \And
    Odellia Boni\\ {\footnotesize odelliab@il.ibm.com} 
   \AND
    Yannis Katsis\\ {\footnotesize yannis.katsis@ibm.com} \And 
    Sara Rosenthal \\ {\footnotesize sjrosenthal@us.ibm.com} \And 
    Benjamin  Sznajder \\ {\footnotesize benjams@il.ibm.com} \And 
    Marina Danilevsky\\ {\footnotesize mdanile@us.ibm.com}
  \AND
  IBM Research - AI 
  }

\begin{document}
\maketitle
\begin{abstract}

Large Language Models (LLM) have become a popular approach for implementing Retrieval-Augmented Generation (RAG) systems, and a significant amount of effort has been spent on building good models and metrics. In spite of increased recognition of the need for rigorous evaluation of RAG systems, few tools exist that go beyond the creation of model output and automatic calculation.
We present \system{}, an introspection platform for performing a comprehensive analysis of the quality of RAG system
output. \system{} allows the user to analyze aggregate and instance-level performance of RAG systems, using both human and algorithmic metrics as well as annotator quality. \system{} is suitable for multiple use cases and is available publicly to the community{$^1$}. A \textbf{live instance} of the platform is available at \url{https://ibm.biz/InspectorRAGet}.

\end{abstract}

% What problem does the proposed system address?
% Why is the system important and what is its impact?
% What is the novelty in the approach/technology on which this system is based?
% Who is the target audience?
% How does the system work?
% How does it compare with existing systems?
% How is the system licensed?
% How was the system evaluated? Were user studies/human evaluation experiments conducted?

\section{Introduction}

\begin{figure*}[t]
    \centering
    \includegraphics[width=\textwidth]{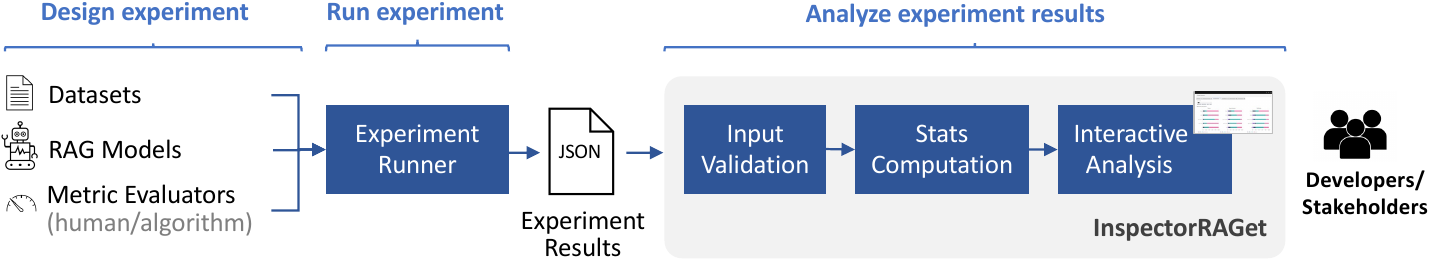}
    \caption{RAG evaluation life cycle. Evaluations of the RAG output are analyzed using \system.}
    \label{fig:architecture}
\end{figure*}

The recent advances in Large Language Models (LLMs) have led to an explosion of research on Retrieval-Augmented Generation (RAG): combining generative LLMs with data retrieval to provide responses grounded on authoritative document collections \cite{rag-neurips-2020}. 
RAG systems have been deployed in diverse domains (see \cite{gao2024retrievalaugmented} for a recent survey). 

There has been increased recognition of the importance of RAG evaluation \cite{longpre2024responsiblefoundationmodeldevelopment}. This consists of 1) designing, 2) running, and 3) analyzing experiments (see~Section \ref{sec:life-cycle}). 
 The research community has mainly limited analyzing to aggregate metrics via
%to support leaderboard-style aggregate performance evaluation
 benchmark datasets \cite{liu2023recall,chen2023benchmarking}, evaluation metrics \cite{es2023ragas} as well as evaluation frameworks (such as RAGAs \cite{es2023ragas} and ARES \cite{saadfalcon2023ares}).

For any input (e.g., a question), a RAG system runs a retriever and passes the input and retrieved passages to a generator.
%to get \textit{RAG output}. 
This \textit{RAG output} can then be evaluated on a variety of metrics. The output is affected by variations in retriever and generator models as well as different model configurations.
%, which entails specific requirements on analysis.
Improving system performance requires actionable analysis of these variations, specific to RAG.
%The ability to seamlessly analyze these variations, on both the aggregate and instance level to understand the role of each input. 
We present a platform for performing a comprehensive analysis of the quality of \textit{RAG output}:
%that includes:

\textbf{Holistic analysis:} End-to-end analysis of RAG system performance involves aggregate evaluations along several dimensions and configurations, enabling continuous benchmarking of models and datasets \cite{gehrmann-etal-2022-gemv2}. A comprehensive set of RAG evaluation metrics comprises algorithmic scores, LLM judges \cite{es2023ragas} and human judgements, and our platform enables comparison and correlation analysis to create a full picture of performance.

%This also enables an analysis and comparison of metrics whether algorithmic scores, LLM judges \cite{es2023ragas}, and human judgements geared for RAG experiments. 

%\textbf{Quality Analysis:} A thorough analysis of \textit{annotator behavior} is required to identify and improve ambiguous guidelines, complex data points, and underperforming annotators. Furthermore, the effectiveness of the evaluation metrics at representing the desired model RAG functionality (e.g., faithfulness to context) should be interrogated. Finally, the dataset should be thoroughly inspected during evaluation, particularly RAG-specific \textit{dataset properties} such as corpus information, relevant passages, and question attributes. By enabling these functionalities, we provide much needed context for the observed quantitative results.
%
% REWRITING SUGGESTION (YANNIS)
%\textbf{Quality analysis of all aspects of a RAG experiment:} While most evaluation works focus on the quality of model outputs, it is equally important to analyze the quality of all other aspects of a RAG experiment. These include analysis of (a) the annotator behavior, to identify/improve ambiguous guidelines and underperforming annotators, (b) the effectiveness of the evaluation metrics at modeling the desired functionality (e.g., faithfulness to context), and (c) the dataset on which models are evaluated. This enables improving the experiment as well as providing much needed context for the intepretation of the observed results. 

\textbf{Quality analysis of all aspects of a RAG experiment:} While most evaluation works focus on the quality of model outputs, it is important to also analyze the quality of all other aspects of a RAG experiment. These include (a) the annotator behavior (e.g. inter-annotator agreement and underperforming annotators) %to identify/improve ambiguous guidelines and underperforming annotators, 
(b) the effectiveness of evaluation metrics at modeling the desired functionality (e.g., faithfulness to context(s)), and (c) the RAG-specific properties of the employed dataset (e.g., relevant passages and question attributes). This analysis can be used to improve the experiment as well as providing context for the interpretation of the results.

%\textbf{Quality Analysis:} It also is important to have a comprehensive analysis of the quality of the actual evaluators (human or automatic), the chosen metrics, and characteristics of the data. The platform also allows a thorough analysis of \textit{annotator behavior} for human judgements. This enables researchers to identify and improve ambiguous guidelines, complex data points, and underperforming annotators, resulting in higher quality human evaluations. The dataset is also subjected to a thorough inspection during evaluation to understand its distribution and \textit{dataset properties} including RAG specific information such as corpus information, relevant passages, and question properties. This provides much needed context for the observed quantitative results.

%\textbf{Error Analysis:} Actionable error analysis for RAG requires understanding the relationships between the question, response, passages, and evaluation metrics (human and automatic). Our instance-level visualization enables the developer to identify the source of undesirable output, correct erroneous reference answers, clarify ambiguous instances, and form hypotheses for model improvement.

\textbf{Error analysis through instance-level inspection:} Actionable error analysis for RAG requires going beyond the typical aggregate-level statistics into inspection of individual instances. This is essential for developers to identify the source of undesirable output, correct erroneous reference answers, clarify ambiguous instances, and form hypotheses for model improvement.

%\textbf{Error Analysis:} It is also necessary to enable the user to perform actionable error analysis. We support the detecting and inspecting of related \textit{individual} instances. This visualizes the input of questions with passages and generated RAG output along with all the metrics in one view. This enables identifying the source of undesirable output, erroneous reference answers, and clarifying ambiguous instances. 

%played by the input data, retriever, generator, and all together

% It also is important to have a comprehensive analysis of quality of the actual evaluators (human or automatic), quality of the chosen metrics, characteristics of the data

% [[[[ commenting this out but may be some good notes

% Finally, these insights are actionable, they allow monitoring over time and enable adjustments to the system across all aspects - data, evaluation, and models.

%  we should try and tie into figure 1 and perhaps update the figure as well.

%  tie into earlier where we say there is much work on models/metrics/data but the ability to understand, analyze, and compare is non-trivial.

%  Our platform is geared specifically for RAG systems where the expectation is a question or conversation and passages are provided as input and an answer is provided as output.

%  ]]]

 %However, when adopting RAG systems in practice, users must continuously analyze both overall performance and instance-level outputs for their specific use cases, compare architectures, understand shortcomings, make targeted adjustments and decide when to deploy. 

 Currently, these analyses can only be done piecemeal, requiring the manual examination of evaluation output using various ad-hoc data processing scripts and spreadsheet tools which are not sustainable, lack re-usability, and are difficult to interpret.
% Our goal is to significantly lower the required effort for this important and undersupported aspect of RAG system development. 
We provide the first platform with end-to-end capabilities for thorough RAG analysis, an undersupported aspect of RAG system development intended for researchers, model developers, and stakeholders. Our contributions are as follows:

\begin{enumerate}[leftmargin=*]
 \setlength\itemsep{.05em}
\item We present \system, a rich interactive platform for performing a comprehensive analysis of RAG system output quality.
%: performance benchmarking, aggregate and instance level analysis, a holistic view of results via a mix of metrics, annotator qualification, and dataset characterization.
\item We evaluate \system\ on its ability to yield concrete and actionable insights on two use cases, on both new and existing datasets.
\item We open source the platform to the community\footnote{\url{https://github.com/IBM/InspectorRAGet}} and host it on HuggingFace\footnote{\url{https://ibm.biz/InspectorRAGet}}.
\end{enumerate}

% Potential references:
% Analyzing/critiquing metrics for telecom domain: https://arxiv.org/pdf/2407.12873

\section{The RAG Evaluation Life Cycle}
\label{sec:life-cycle}

\begin{figure*}[t]
    \centering
    \includegraphics[width=\textwidth]{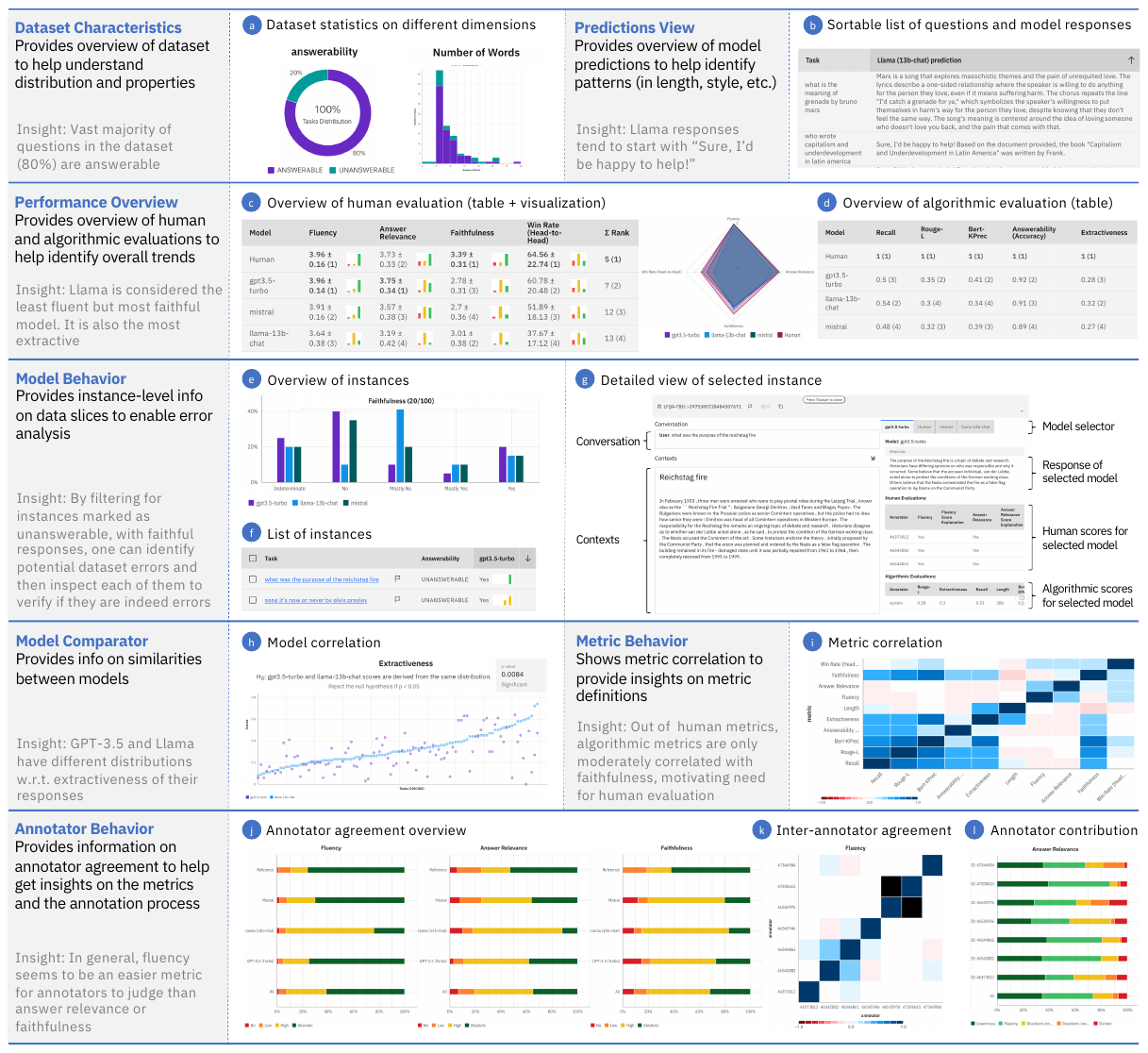}
    \caption{Illustration of \system's core views and corresponding visualizations. Screenshots are drawn from the RAG model performance use case, described in Section \ref{sec:usecase1}.}
    \label{fig:combo.interface}
\end{figure*}

%We begin by briefly describing the life cycle of evaluating RAG systems, and situating \system\ within it, as shown in Figure \ref{fig:architecture}.
We briefly describe the life cycle of evaluating RAG systems, and situating \system\ within it, as shown in Figure \ref{fig:architecture}.
Evaluating RAG systems involves three main steps: %\\
\vspace*{0.3cm}

%\noindent {\bf (1) Design the evaluation experiment:} One starts by designing the \emph{evaluation experiment} (or simply \emph{experiment}), which consists of\ {\it Models, Datasets, Metrics} and {\it Metric Evaluators}, as defined below.

%\noindent {\bf (2) Run the evaluation experiment:} Once the experiment has been designed, the next step is to compute the evaluation scores. This involves (a) generating the model responses/output and (b) passing the responses to the metric evaluators (algorithms or humans) to produce instance-level scores.

%\noindent {\bf (3) Analyze the experiment results:} Finally, the experiment results need to be analyzed to gain actionable insights about the models, datasets, metrics, annotation quality, etc.

\noindent {\bf (1) Design the evaluation experiment}, which consists of\ {\it Models, Datasets, Metrics} and {\it Metric Evaluators}, as defined below.

\noindent {\bf (2) Run the evaluation experiment} and compute the evaluation scores. This involves (a) generating the model responses/output and (b) passing the responses to the metric evaluators (algorithms or humans) to produce instance-level scores.

\noindent {\bf (3) Analyze the experiment results} to gain actionable insights about the models, datasets, metrics, annotation quality, etc.

\subsection {\system\ for RAG Evaluation}

\system\ focuses on the third step of the RAG evaluation life cycle. To analyze an experiment using the platform, users upload the standardized experiment results JSON file which contains the following information:

\noindent \textbullet\ {\it Datasets:} 
The set of data instances included in the experiment, each containing user input (e.g., question/conversation), contexts (e.g., passages),
and, optionally, reference responses.

\noindent \textbullet\ {\it Model metadata:} Name and description of the RAG models that were evaluated.

\noindent \textbullet\ {\it Metric metadata:} Metadata about the \emph{metrics} on which model responses were evaluated. These include metric name,  type (i.e., \emph{algorithmic}, LLM-as-a-judge and \emph{human}, such as crowd workers), and scale (yes/no, Likert scale, numeric, etc.).

\noindent\textbullet\ {\it Model output and Evaluation scores:} The model responses and evaluation scores for the models along each metric on every data instance~\footnote{In case of multiple annotators per metric, such as when multiple humans annotate a single response, multiple evaluation scores are included: one per annotator.}.
\vspace{0.3cm}

The platform and the experiment results file format have been designed in a model-agnostic and metric-agnostic way to support analysis of diverse evaluation experiments (see Appendix \ref{sec:input-format}).

Once provided with the input file, \system\ validates it for errors (e.g., missing evaluation scores) and augments it with additional statistics (e.g. inter-annotator agreement for human evaluations). The augmented results power the platform's frontend: a visual analytics application that enables users to interactively analyze and gain insights on different aspects of the experiment.%, as described in the next section.

%Once provided with the input file, \system\ validates it for errors (e.g., missing evaluation scores) and augments it with additional statistics (e.g. inter-annotator agreement for human evaluations). The augmented experiment results are then used to power the platform's frontend; a visual analytics application that enables users to interactively analyze and gain insights on different aspects of the experiment, as described in the next section.

Finally, to help users employ the platform as part of the broader RAG evaluation life cycle we also provide sample experiment runners, showing how to run experiments using popular evaluation frameworks and output the results in the format expected by \system\ for analysis of experiment results. Our experiment runners showcase integrations with the Language Model Evaluation Harness~\cite{lm-eval-harness} and RAGAs~\cite{es2023ragas} evaluation frameworks, as well as HuggingFace~\cite{wolf2020huggingfaces} (see Appendix \ref{sec:notebooks}).

%assets~\cite{wolf2020huggingfaces} (e.g. datasets, models, and metric evaluators) and output the results in the format expected by \system\ (see Appendix \ref{sec:notebooks} for details).

%Finally, while \system\ focuses on the analysis of the experiment results, to help users employ the platform as part of the broader RAG evaluation life cycle we have also implemented a sample experiment runner in a Python notebook, showing how to run experiments using HuggingFace assets~\cite{wolf2020huggingfaces} (e.g. datasets, models, and metric evaluators) and output the results in the format expected by \system\ (see Appendix \ref{sec:notebooks} for details).

\section{ \system\ Platform}
\label{sec:platform}

\system{} is a React web application built with NextJS 13 framework\footnote{\url{https://react.dev}, \url{https://nextjs.org}}. We use the Carbon Design System\footnote{\url{https://carbondesignsystem.com}} for the user interface. The experiment results are  provided as a json input file that is loaded on the platform. This enables our platform to be lightweight; it can easily be run on virtual machines or even personal laptops/desktops. To enable privacy, \system{} is a stateless application and does not retain any uploaded datasets. \system{}'s frontend  offers a series of views (presented as separate tabs), each tailored to a different aspect of the analysis process. %(see Appendix \ref{sec:implementation} for implementation details). 
 Excerpts of these views are shown in Figure \ref{fig:combo.interface}. We describe each view along with hypotheses that can be formed, analyses it enables, and insights which may require further investigation.
 %In Section \ref{sec:insights}, we then demonstrate how these views can be used to extract actionable insights for two different use cases.   

\subsection{Dataset Characteristics}

The Dataset Characteristics view (Figure \screenshot{a}) provides details regarding characteristics of the dataset such as answerability (e.g. answerable, unanswerable, partial), question type (e.g. factoid, comparative, explanation), and question length. These details can provide a quick snapshot of dataset trends, outliers, and potential biases.

\subsection{Predictions Table}
When analyzing an experiment, it helps to first see a few examples of instances.
%To this end, 
\system{}'s predictions view shows a table of all questions with the respective model responses (Figure \screenshot{b}). This view not only gets developers acquainted with the experiment but also helps them spot patterns in model responses (e.g., length, repetitions of text).

\subsection{Performance Overview}
After getting acquainted with the experiment, developers and stakeholders can get an overview of the experiment results through the overall performance view. This view shows the aggregate score and ranking of each model for each evaluation metric. This information is rendered both in tabular form, as well as through a Radar chart with separate tables/visualizations for human and algorithmic metrics (Figures \screenshot{c} and \screenshot{d}).

While designing this view, special attention was given to quantifying the uncertainty present in human evaluations to avoid misinterpretation of the results. Aggregate evaluation scores for human metrics are shown along with (a) their standard deviation and (b) a visualization of inter-annotator agreement (through sparkline charts).
Despite its resemblance to leaderboards \cite{NEURIPS2023_91f18a12MTBench, hendrycks2021measuringMMLU}, the goal of this view is not only to declare winners but to also make initial observations that need to be further explored.
 
\subsection{Model Behavior}
\label{sec:model-behavior}
After obtaining an overview of aggregate model performance, one can drill down and inspect model responses for individual instances through the model behavior view. Users start the analysis by filtering instances based on criteria, such as dataset domain, whether a question is answerable, etc. The view then shows a histogram of model scores for all instances satisfying the filter (Figure \screenshot{e}), as well as a sortable table of the actual instances (Figure \screenshot{f}). Upon selecting an instance, users see a detailed view of the instance, including the conversation/question, context(s), the model responses and their respective evaluation scores (Figure \screenshot{g}). Instance-level analysis is crucial for conducting error analysis and understanding the root causes of issues observed in other views, as well as identifying ``I know it when I see it" types of issues. 
The view also provides users with the functionality to easily copy, flag or comment on instances. 

\subsection{Model Comparator}
In addition to inspecting individual model responses, one can also analyze entire models. This is enabled by the model comparator view, which for a chosen pair of models and a metric, shows whether the scores of the two models for the selected metric are derived from the same distribution (Figure \screenshot{h}). This is shown both through a scatter plot, depicting the scores of the two models for the metric, as well as through the result of a statistical significance test, computed using Fisher's randomization method \citep{10.1145/1321440.1321528}.
Continuing the support for instance-level analysis, the view also allows one to drill down and inspect the instances where two models received very similar/dissimilar evaluation scores. This view can be useful for spotting similarities and differences between models.     

\subsection{Metric Behavior}
Similarly to comparing models, one can also compare metrics. This is accomplished through the metric behavior view, which shows the Spearman correlation scores for each pair of metrics. This allows developers to gain insights on metrics, such as identifying whether an automatic metric correlates well with human judgements or whether supposedly orthogonal metrics correlate with each other (see fluency and answer relevance in Figure \screenshot{i}), hinting at issues with metric definitions.

\subsection{Annotator Behavior}
Finally, when human evaluation is performed, it is imperative to also analyze its quality. This is enabled by the annotator behavior view, which contains 
two types of visualizations related to annotation quality: (a) \emph{Model-level visualizations} that show the annotator agreement when evaluating each model (Figure \screenshot{j}). These can help provide insights on challenges that certain models pose to human annotators (e.g., if humans had trouble reaching agreement when evaluating certain models).
(b) \emph{Annotator-level visualizations} depicting individual annotators' performance. These include a visualization of inter-annotator agreement (Figure \screenshot{k}), computed using Cohen's kappa \cite{cohen1960coefficient}), annotator contribution (i.e., how often an annotator agreed with the majority) (Figure \screenshot{l}), and annotation time (not shown for space reasons).
Insights drawn from this view can drive additional analyses (e.g., to understand why annotators had trouble annotating the responses of certain models) or changes to the experiment setting (e.g., give feedback to individual annotators or improve the annotation guidelines).

\newcommand\sbullet[1][.5]{\mathbin{\vcenter{\hbox{\scalebox{#1}{$\bullet$}}}}}

\begin{table*}[t!]
\small
\centering
%\begin{tabular}{@{}p{1.5cm}lp{6.2cm}p{5.5cm}@{}}
\begin{tabular}{@{}p{1.7cm}p{1cm}p{6.7cm}p{5.3cm}@{}}
\toprule
\textbf{Use Case} & \textbf{Dataset} & \textbf{Models} & \textbf{Evaluation Metrics}\\
\midrule
RAG Model Performance & \datasetname{} & Llama-13B~\cite{touvron2023llama}, GPT-3.5 Turbo~\cite{brown2020languageGPT}, Mistral~\cite{jiang2023mistral} & $\sbullet[.75]$ Human: fluency, answer relevance, faithfulness, win-rate \newline $\sbullet[.75]$ Algorithmic: 
Recall, Rouge, Bert-KPrec, Answerability, Extractiveness, Length \\
\midrule
LLM-as-a-Judge Performance & MT-Bench & Alpaca-13B~\cite{alpaca}, Claude-v1~\cite{claude}, GPT-3.5~\cite{brown2020languageGPT}, GPT-4~\cite{openai2024gpt4}, Vicuna-13B~\cite{vicuna2023} & $\sbullet[.75]$ Human: Win-Rate \newline  $\sbullet[.75]$ Algorithmic: LLM-Judge GPT-4 \\
\bottomrule
\end{tabular}
\caption{Evaluation settings for the use cases presented in this paper.}
\label{tab:use-cases}
\end{table*}

\section{Evaluating \system{}}
\label{sec:insights}

To showcase the value of \system{}, we evaluate its ability to yield actionable insights on two use cases: (1) \emph{Analyzing RAG Model Performance:} We collected human judgements of model responses for a RAG dataset. This allows us to show the full scope of our platform for comparing human and algorithmic metrics as well as multiple annotators. (2) \emph{Analyzing LLM-as-a-Judge Performance:} We use annotations comparing human and LLM-as-a-judge on model output for multi-turn question answering. 

The experiment setting for each use case\footnote{The use cases are at  \href{https://ibm.biz/InspectorRAGet}{ibm.biz/InspectorRAGet}} (e.g. dataset, models, and evaluation metrics) is summarized in Table~\ref{tab:use-cases}. For each use case we describe the discovered insights along with the \textit{Source} views used to identify them and propose possible \textbf{Actions} for improving the RAG experience.
%We provide platform screenshots of the findings in Appendix~\ref{app:screenshots}.

\subsection{Analyzing RAG Model Performance}
\label{sec:usecase1}

To illustrate the full capabilities of our platform on RAG we performed our own manual evaluation on \datasetname{}~\cite{rosenthal2024clapnq}, a long form question answering dataset. We also explored Fine-Grained ASQA~\cite{stelmakh-etal-2022-asqa, DBLP:journals/corr/abs-2306-01693_finegrained_asqa} and InstructQA~\cite{adlakha2023evaluating} as alternative RAG model evaluations but both of these evaluations did not provide enough details suitable for analysis (e.g., missing model information and incomplete human annotations). 

\datasetname{} is built on the portion of the Natural Questions dataset \cite{kwiatkowski-etal-2019-natural} that only has a long answer (gold passage) without an extractive short answer. The responses in \datasetname{} are grounded on the gold passage and must be concise and complete. We ensure that every question is evaluated by the same algorithmic metrics and the same number of human evaluators (3 per question). The human evaluation annotation task was completed on Appen\footnote{\url{https://www.appen.com/}}, a crowdsourcing platform for collecting high-quality annotations. Each task has a question, grounding passage, and multiple randomly shown model responses for the annotator to provide their evaluations. We asked annotators to evaluate the answers on three metrics: \textit{fluency, answer relevance}, and \textit{faithfulness} as commonly used in the literature ~\cite{es2023ragas, chiang-lee-2023-large}. In addition, we also asked them to perform a head-to-head comparison of all models for \textit{win-rate}. These annotations allow us to use the platform to compare and draw insights from the models, data, metrics, and annotators. We next describe key insights we found using the platform.

\textbf{Low Performance:} It is clear that Llama is the model least preferred by the annotators based on win-rate. Its answers are somewhat faithful but not relevant. It is also the only one that has lower fluency for 29/100 tasks. Annotators did not like the phrase “Sure I’d be happy to help” which was used frequently by Llama. Llama answers are also the longest and most extractive (which correlates with high faithfulness). Source: \textit{Predictions, Performance Overview, Model Behavior, Model Comparator} \textbf{Action:} Propose to stakeholders to reconsider Llama as model of choice for this use case.

\textbf{Algorithmic vs Human:} Based on algorithmic metrics Mistral is the worst, but it was considerably preferred over Llama by human evaluators. We suspect that this is because its responses are slightly shorter, which biases Recall. Source: \textit{Performance Overview}. \textbf{Action:} Continue including human evaluation in future evaluation rounds, as it provides different insights than algorithmic metrics.

\textbf{Dataset Inconsistencies:}  During the human evaluation the reference responses (labeled Reference) were rated highest by the annotators which shows that the dataset is of good quality. However, the annotators disagree most on these responses. Filtering to see specific instances of disagreement, shows that the most disagreement is 25 cases of Answer relevance. Opening up an %indeterminate
example where there was no agreement, shows a case of a  tricky question+answer which explains the disagreement. Source: \textit{Performance Overview}, \textit{Annotator Behavior}, \textit{Model Behavior}. \textbf{Action:} Investigate whether data inconsistencies should be improved or removed, and if further annotator training is needed. 
%Propose revisiting the experiment design via data analysis and improvement.

\textbf{Data Types:} It is also interesting to explore the answerable and unanswerable splits separately. Examples of insights include finding examples where an unanswerable question was answered by the models and is faithful, which may indicate that the question is actually answerable. One may also search for unanswerable examples that are not faithful to show that the model is coming up with an answer from its own knowledge or is hallucinating. Source: \textit{Performance Overview}, \textit{Model Behavior}, \textit{Dataset Characteristics}. \textbf{Action:} Performance can differ due to question type, domain and other characteristics. Provide model feedback to stakeholders per dimension.

\textbf{Metric correlations:} Win rate is correlated mostly with human metrics. This highlights the importance of human evaluation. The algorithmic metrics do not indicate which responses are preferred. Extractiveness and BertK-Precision is correlated with faithfulness. This is expected as a faithful model will have information extracted from the passage either directly or reworded. The metric correlation can be used to inspect why there are cases with low extractiveness/BertK-Precision and high faithfulness. This may highlight annotator confusion. Source: \textit{Metric Behavior}. \textbf{Action}: Perform a human evaluation for accurate model preference. Investigate possible annotator confusion.

\subsection{Analyzing LLM-as-a-Judge Performance}

%There has been a significant uptick in the popularity of LLMs as judges to evaluate the quality of LLM responses in RAG settings, complemented by active research in the efficacy of these judges \cite{chiang-lee-2023-large,shen-etal-2023-large}. Our platform also seamlessly supports analyzing the performance of LLM-as-a-judge approaches, and we illustrate this use case using MT-Bench. MT-Bench~\cite{NEURIPS2023_91f18a12MTBench} is a multi-turn question answering dataset of 80 high quality multi-turn questions. 
%The paper introducing this dataset explores using LLM-as-a-judge as a means of evaluating benchmarks by comparing it to human evaluation by 58 expert annotators. To this end, they release human and algorithmic GPT-4 judgements on the MT-bench dataset. 

There has been a significant uptick in the popularity of LLMs as judges to evaluate the quality of LLM responses in RAG settings, complemented by active research in the efficacy of these judges \cite{chiang-lee-2023-large,shen-etal-2023-large}. Our platform also seamlessly supports analyzing the performance of LLM-as-a-judge approaches, and we illustrate this use case using MT-Bench~\cite{NEURIPS2023_91f18a12MTBench}, a dataset of 80 high quality multi-turn questions. 
The MT-Bench authors compare LLM-as-a-judge with humans by releasing human judgments (provided by 58 experts) and algorithmic GPT-4 judgments on the MT-Bench dataset.
%The MT-Bench authors explore using LLM-as-a-judge to evaluate benchmarks and compare with human evaluation by 58 expert annotators. They release human and algorithmic GPT-4 judgements on the MT-Bench dataset. 

This \system{} use case is not centered around comparing model performance, but rather judge performance. 
The MT-Bench authors discuss several limitations of LLM-as-a-judge: positional bias, verbosity bias, self-enhancement bias (judge favors its own model's answers), and limited reasoning ability (low reasoning and math capability). They show that GPT-4 judge matches human evaluation at over 80\%. We expand on their insights and introduce additional ones.

\textbf{Self-Enhancement Bias:} The expert annotators tend to agree with each other and match GPT-4 closely on all models, but GPT-4 seems to prefer responses from its own model. Source: \textit{Annotator Behavior, Model Behavior}. \textbf{Action:} Inform stakeholders of the slight bias of GPT-4.

\textbf{Verbosity Bias:} Answer length is strongly correlated with win-rate. This is true for LLM-as-a-judge and human annotators. Claude-v1, GPT-3.5, and GPT-4 usually have the longer response and win, while Llama and Alpaca rarely have the longer response and usually lose. Source: \textit{Model Behavior}, \textbf{Action}: Analyze data and share with stakeholders. Do the answers need to be long for these questions? What is missing in the short responses?

\textbf{Positional Bias:} As hinted in the paper, the first answer being favored is not as much of an issue for GPT-4. We believe it also may be model dependent. For instance Llama is always shown to GPT-4 first but also almost always loses and Claude-v1 is always shown second and always wins. Source: \textit{Model Behavior}, \textbf{Action}:  Revisit evaluation design. Investigate if there may be some unintentional bias because these two models were never randomized.

\section{Conclusion}

We present \system{}, a publicly available\footnote{\url{https://github.com/IBM/InspectorRAGet}} platform to empower researchers, developers, and stakeholders to gain a deeper understanding of the strengths and limitations of RAG systems. It includes aggregate-level and instance-level views as well as capabilities to explore human and algorithmic metrics and annotator behavior, allowing for a more holistic analysis.
%We evaluate \system{} on its ability to
Our evaluation shows \system{}'s ability to yield concrete and actionable insights on two pertinent use cases - analyzing RAG model performance and LLM-as-a-judge performance. 
%that showcase and highlight the value of these features, 
We publicly release our platform, input files used in this paper, and notebooks. %In the future, we will explore adding additional capabilities to \system{} to facilitate comprehensive pattern discovery, as well as extending its analytical capabilities beyond RAG to other popular LLM tasks, such as summarization and code generation. 
In the future, we will explore augmenting \system{} to facilitate comprehensive pattern discovery, as well as extending its analytical capabilities beyond RAG to other popular LLM tasks, such as summarization and code generation.

\section{Ethical Considerations}

A key claim in our paper is to include human annotations in the evaluation process. Human evaluation is subjective and prone to errors. We expect even the best annotators to make mistakes. We suggest mitigating this by including multiple annotators per question, but biases still may occur.

All proposed actions are based on evidence provided by the tool and considered to be our own opinions for possible areas of improvement. Any biases in the datasets may impact the analysis. The platform is meant to be used as a means of drawing conclusions and insights of RAG systems; it does not create its own conclusions or insights.

We acknowledge that there are accessibility limitations in the platform and we plan on providing additional features to improve these limitations.

\bibliography{anthology, custom}

\appendix

\newpage

\section{Experiment Results File Format}
\label{sec:input-format}
As \system\ is a web application, we naturally gravitated towards adopting JSON as the input format. Our prescribed structure for an experiment results file is intuitive and strives to minimize repetition of information. 

The experiment result file can be broadly split into six sections along their functional boundaries. The first section captures general details about the \emph{experiment}, including its name, description, and timestamp. The second and third sections describe the sets of \emph{models} and \emph{metrics} used in the experiment, respectively. The fourth and fifth sections cover the dataset, in the form of a list of \emph{documents/passages} that are used in the experiment and a list of \emph{tasks}, each representing a data instance, which is composed of the user input (e.g., question/conversation), references to the corresponding documents/passages, and, optionally, a list of reference responses. Finally, the sixth section includes information about the \emph{outcome of the evaluation}, in the form of the scores of the different evaluation metrics for each task.

Note, that as part of the experiment results file, one is not providing the implementation of the metrics and models used, but rather high-level \emph{metadata} about them (e.g., the name of a metric and its scale - e.g., yes/no, numeric - and the name of a model) along with their \emph{outputs} (i.e., the resulting evaluation scores and model responses). This separation of the  evaluation experiment implementation and run from the analysis of the results, allows \system{} to be agnostic of the specific models or metrics used, thus allowing the analysis of diverse evaluation experiments.   

%Figure \ref{fig:input-format} describes the requirements associated with each section in detail.
A detailed description of the input format can be found on our GitHub repository~\footnote{\url{https://github.com/IBM/InspectorRAGet}}.  
%\label{sec:input-format}
%\begin{figure*}
%    \centering
%    \includegraphics[width=0.8\textwidth]{images/analytics_platform_input_file_format.png}
%    \caption{Format of the Experiments Results input file}
%    \label{fig:input-format}
%\end{figure*}

\section{Sample Experiment Runners}
\label{sec:notebooks}
To help users employ \system\ as part of the broader RAG evaluation life cycle, we have also released sample experiment runners, showcasing how to use \system\ in combination with popular evaluation frameworks. Each experiment runner is provided in the form of a Python notebook, which demonstrates how to use the corresponding evaluation framework to run an evaluation experiment and transform its output to the input format expected by \system\ for an analysis of the evaluation results. As of this writing, we have released notebooks demonstrating integrations of our platform with the following popular frameworks:
\begin{itemize}
\item \emph{Language Model Evaluation Harness}~\footnote{\url{https://github.com/EleutherAI/lm-evaluation-harness}}~\cite{lm-eval-harness}; a popular evaluation framework used to evaluate LLMs on different tasks.
\item \emph{RAGAs}~\footnote{\url{https://github.com/explodinggradients/ragas}}~\cite{es2023ragas}; a popular evaluation framework specifically designed for the evaluation of RAG systems through LLM-as-a-judge techniques.
\item \emph{HuggingFace}~\footnote{\url{https://huggingface.co}}~\cite{wolf2020huggingfaces}, which offers libraries and assets (incl. datasets, models, and metric evaluators) that can be used to both create and evaluate RAG systems.
\end{itemize}
All experiment runners are available on the platform's GitHub repository.

\end{document}